%% file: main.tex
\documentclass[twocolumn]{aastex63}
\pdfoutput=1 
\usepackage{amsmath,amstext,amssymb}
\usepackage[T1]{fontenc}
\usepackage{apjfonts} 
\usepackage{subfigure}

\usepackage{xcolor}
\usepackage{soul}
\usepackage{comment}
\usepackage{graphicx}

\graphicspath{{figures/}}

\newcommand{\de}{{\rm d}}

\shorttitle{Probing gravity and growth of structure with GWs and galaxies' peculiar velocity}
\shortauthors{A.~Palmese \& A.~Kim}

\begin{document}

\title{Probing gravity and growth of structure with gravitational waves and galaxies' peculiar velocity
}

\author{A.~Palmese}
\affiliation{Fermi National Accelerator Laboratory, P. O. Box 500, Batavia, IL 60510, USA} 
\affiliation{Kavli Institute for Cosmological Physics, University of Chicago, Chicago, IL 60637, USA}
\email{palmese@fnal.gov}

\author{A.~G.~Kim}
\affiliation{Lawrence Berkeley National Laboratory, Berkeley, CA 94720, USA}

\begin{abstract}
The low-redshift velocity field is a unique probe of the growth of cosmic structure and gravity. We propose to use distances from gravitational wave (GW) detections, in conjunction with the redshifts of their host galaxies from wide field spectroscopic surveys (e.g. DESI, 4MOST, TAIPAN), to measure peculiar motions within the local Universe. Such measurement has the potential to constrain the growth rate $f\sigma_8$ and test gravity through determination of the gravitational growth index $\gamma$, complementing constraints from other peculiar velocity measurements. We find that binary neutron star mergers with associated counterpart at $z\lesssim 0.2$ that will be detected by the Einstein Telescope (ET)
will be able to constrain $f\sigma_8$ to $\sim 3\%$ precision after $10$ years of operations when combined with galaxy overdensities from DESI and TAIPAN. If a larger network of third generation GW detectors is available (e.g. including the Cosmic Explorer), the same constraints can be reached over a shorter timescale ($\sim 5$ years for a 3 detectors network). The same events (plus information from their hosts' redshifts) can constrain $\gamma$ to $\sigma_\gamma\lesssim 0.04$. 
This constraint is precise enough to discern General Relativity from other popular gravity models at $3\sigma$. This constraint is improved to $\sigma_\gamma\sim 0.02-0.03$ when combined with galaxy overdensities. The potential of combining galaxies' peculiar velocities with gravitational wave detections for cosmology highlights the need for extensive optical to near--infrared follow--up of nearby gravitational wave events, or exquisite GW localization, in the next decade. 
\end{abstract}


\reportnum{FERMILAB-PUB-20-185-AE}

\section{Introduction}

The motion of galaxies on top of the cosmological expansion, i.e.\ their peculiar velocity field, follows the inhomogeneous clustering of structure in the Universe and the laws of gravity. Peculiar velocities of galaxies thus encode important information about large scale structure and its growth, and can probe models of gravity.
The velocity field can be studied in several ways. One possibility is through redshift space distorsions (RSD), since peculiar motions alter the correlations between galaxies along the line of sight. Another option is to derive peculiar velocity measurements from redshift surveys using galaxies that also have a distance estimate, so that the contribution due to the cosmological expansion can be subtracted out. Such distances can be estimated from the fundamental plane relation for elliptical galaxies, the Tully--Fisher relation for spiral galaxies, and, as it has been suggested more recently, from Type~Ia Supernovae (SNe~Ia) (e.g.\  \citealt{gordon,abate,huterer,kim_wp,kim2020}).

Gravitational wave (GW) detections also provide a distance measurement. If a single galaxy can be associated with a GW event, either through an observable electromagnetic (EM) counterpart, or thanks to an exquisite localization by the GW detectors, the host galaxy will also inherit the GW distance measurement. The first cosmological parameter estimates from GW detections (\citealt{2017Natur.551...85A}; \citealt*{darksiren1}) rely on this principle, and are performed using the ``standard siren'' method \citep{schutz}, which makes use of the distance--redshift relation to probe the expansion of the Universe. In other words, the GW distance estimates are used with host galaxies' redshifts to populate the Hubble diagram. In these works, the peculiar motion of galaxies can give rise to systematic uncertainties and biases, and several works have studied how these can be properly taken into account in standard siren measurements  \citep{howlett_davis,Mukherjee,nicolaou2019impact}. However, peculiar motions also contain interesting cosmological information that we could exploit.

\citet{2018PhRvD..98f3503W} recently suggested using gravitational waves with counterparts as distance indicators to measure peculiar velocities, and discussed implications for estimates of $\Lambda$CDM cosmological parameters using a comparison method between the density field of galaxies and peculiar velocities from GWs using measurements up to 190 Mpc. In this work, we extend this method to include auto--correlations of overdensities and of the velocity field, other than cross--correlations, including information from the CMB as explained in more detail in Appendix \ref{app:comparison}.
Moreover, we show how this method is interesting beyond 190 Mpc, and beyond $\Lambda$CDM to determine the gravitational growth index $\gamma$ as a tool to test General Relativity (GR).

We focus on the results that will be obtained using third generation (3G) gravitational wave detectors, such as the Einstein Telescope (ET; \citealt{Punturo_2010}) and the Cosmic Explorer (CE; \citealt{CE}), because current-generation detectors would not produce luminosity distance measurements precise enough
to be competitive for this analysis (currently $\sigma_d>10\%$; \citealt{LIGOprospects}). Moreover the limited distance horizon of the LIGO/Virgo/KAGRA network ($\lesssim 200$ Mpc for binary neutron star mergers; \citealt{LIGOprospects}) results in small number statistics. On the other hand, 3G GW experiments will detect up to millions of compact object binary mergers across the history of the Universe every year, including binary neutron star mergers (BNS) out to $z\sim 2$ and binary black hole mergers out to $z\sim 20$. Detections of BNSs' GW signal in the local Universe will be highly complete over the full sky, within the redshift range we are interested in for our purposes ($z<0.3$). 
Several works have shown how these sources can be used as standard sirens for precision cosmology (e.g. \citealt{2010CQGra..27u5006S,Zhao_2011,Cai_2017,zhao}), although attention has to be paid at the way the analysis is carried out to avoid introducing biases \citep{Keeley_2019,Shafieloo_2020}.

The promise of future ground--based GW detectors is also particularly compelling for peculiar velocity studies since the expected precision on luminosity distance measurements can reach few per cent in the nearby Universe \citep{zhao}, and even sub--percent with a space--based experiment such as DECIGO \citep{decigo,decigo2}. A large number of facilities is expected to be involved in following up these events (e.g. \citealt{chornock2019multimessenger,palmese_WP}), and it is reasonable to assume that the next generation of telescopes will be able to detect optical counterparts to nearby $z<0.3$ events (e.g. LSST, WFIRST; \citealt{Scolnic_2017}). These data can be combined with precise host-galaxy redshifts from ongoing and planned wide--field spectroscopic galaxy surveys, such as the Dark Energy Spectroscopic Instrument (DESI; \citealt{desi}), the 4-metre Multi-Object Spectroscopic Telescope (4MOST; \citealt{4most}) and TAIPAN \citep{taipan}, to derive the constraints presented in this work.

This article is structured as follows. In Section \ref{sec:experiments} we define the specifics of the GW experiment and the galaxy survey that are relevant for the forecast. In Section \ref{sec:method} we present the method used to derive the constraints, including the Fisher formalism. Section \ref{sec:results} contains our forecasted constraints on the gravitational growth index for 3G GW detectors, and a comparison to the expected precision using SNe Ia with peculiar velocities. In Section \ref{sec:conclusion} we present the conclusions. 
Throughout this article, we assume a Flat $\Lambda$CDM cosmology with $H_0=100~ h~{\rm km~s}^{-1}~{\rm Mpc}^{-1}$ and $\Omega_{m0}=0.3.$ Quoted uncertainties are 1$\sigma$.

\section{Experiments specifics}\label{sec:experiments}

In this work, we focus on gravitational wave detections of binary neutron star (BNS) mergers. This is justified by the fact that BNSs are expected to be accompanied by an EM counterpart \citep{Metzger,Barnes_2013}, and this has been observed for the event GW170817 \citep{MMApaper}.
BNS rates are estimated to be 1 to 2 order of magnitude lower than SNe Ia: $\sim 0.25-2.81\times 10^{-6}$ Mpc$^{-3}$ yr$^{-1}$ \citep{190425} versus 
 $2.69 \times 10^{-5}(h/0.70)^3\, \text{Mpc}^{-3} \text{yr}^{-1}$ \cite{2010ApJ...713.1026D}. Note that we use this effective rate for the $z<0.3$ considered in this article, and do not account for the redshift-dependent increase in rest-frame rates.
 While the rates imply that the number statistics of GWs will be less constraining than SNe, the promise of more precise distance measurements in the local Universe and full sky detection sensitivity can still make the method presented here a competitive probe.
The expected distance precision for an ET-CE network is a few percent in the local Universe, depending on the network configuration \citep{zhao}, while
SN standardization methods leave an
$\sim 5\%$ distance uncertainty floor that is
not improved with precision photometry \citep[e.g.][]{2014A&A...568A..22B}.

Moreover, a CE-ET network of multiple detectors can be considered sensitive to all sky directions to first order, in particular for the nearby BNS considered here, resulting effectively in a $4\pi$ experiment. However, we still need to identify EM counterparts, meaning that even if we assume that there is an instrument on the ground and/or in space that is able to identify them at any time, the discovery would still be challenging close to the Galactic plane. We will make the assumption that $\sim 10\%$ of the sky area is lost because of this. We consider different time windows for the experiment, up to 18 years, to show the scaling of the constraints with the number of detected sources.

In addition, we consider perspectives for current generation GW detectors. We take into account a LIGO/Virgo configuration (HLV) at design sensitivity (expected for 2022+), which is able to detect BNS events out to 190 Mpc \citep{LIGOprospects}, with a $\sim 20\%$ uncertainty on the distance on average (e.g. \citealt{Chen_2019}). Several works have shown how independent EM observations can constrain the binary viewing angle \citep{Guidorzi_2017,Hotokezaka,Dobie_2020,dhawan}, thus providing an external constraint that breaks the inclination angle--distance degeneracy, one of the main sources of uncertainty for the distance precision. When estimated independently of the distance, viewing angle constraints can be as good as $\sim 5$ deg for GW170817 when using X-ray and radio data, and $\sim 10$ deg from optical-NIR data \citep{dhawan}, leading to a factor 2 or better improvement on the distance precision. 

In all of the scenarios considered, the EM counterparts need to be accompanied by host galaxy redshifts. While it is realistic to assume that they will be measured as part of the GW follow--up campaigns, it is worth noting that most of these galaxies in the local Universe are likely to already have a spectroscopic redshift measured as part of ongoing and upcoming galaxy surveys such as DESI and TAIPAN. The same surveys will also measure RSDs, and in the following we consider their complementarity with the proposed peculiar velocity surveys. In particular, we consider a galaxy survey with number density of $n_g \sim 10^{-3}~h^{3}~{\rm Mpc}^{-3}$ out to $z<0.2$, and $10^{-4}~h^{3}~{\rm Mpc}^{-3}$ at $0.2<z<0.3$, which is realistic for TAIPAN \citep{Howlett_2017a}, and it is a conservative lower bound for the DESI Bright Galaxy Survey (BGS; \citealt{desi}).

The last ingredient needed in our analysis is the range of scales to consider in the power spectra. Following \citet{2017MNRAS.464.2517H}, the minimum and maximum modes $k$ are set by the linear extent of the GW experiment volume ($k_{\rm min} = (\pi/r_{\rm max}) ~h$/Mpc, with $r_{\rm max}$ being the 
radius of the volume under consideration) and by the smallest scales that are confidently modelled ($k_{\rm max} = 0.1 h$/Mpc), respectively.

\section{Method}\label{sec:method}

The connection between peculiar velocities and the growth of structure is straightforward to understand
in linear theory.  The amplitude
of overdensities scales with
the growth factor $D$, which evolves
as $f \equiv \frac{d\ln{D}}{d\ln{a}}$,
the linear growth rate.
From the conservation of mass,
the velocity field scales
with the overdensity field by
a factor $f$.
The peculiar velocity power spectrum
is related to the overdensity
power spectrum at the time of the CMB as
$P_{vv}\propto (fD \mu)^2P_{\delta \delta}(z=CMB)$.
(See \S\ref{appendix:sec} for an
overview).

The growth of structure depends on gravity. \citet{2005PhRvD..72d3529L} and \citet{2007APh....28..481L} show that the linear growth rate is well--approximated by $f \approx \Omega_{\rm m}^\gamma$ for several gravity models, where $\gamma$ is the growth index. They also find that for General Relativity, $f(R)$, and DGP gravity (see \citealt{HUTERER201523} for a review),
$\gamma=0.55, 0.42, 0.68$, respectively. It is thus interesting to constrain $\gamma$ with a precision of $\sigma_\gamma/\gamma \lesssim 20\%$ at $3\sigma$ (in other words, $3 \sigma_\gamma\sim 0.1$) to be able to discern between GR and other gravity models at $\sim 99\%$ CL.
Using this parameterization, the peculiar velocity power spectrum probes gravity through $\gamma$ in:
\begin{equation}
    fD=a_{\rm CMB} \Omega_{\rm m}^{\gamma}  {\rm e}^{\int_{a_{\rm CMB}}^a \Omega_{\rm m}^{\gamma} \de \ln{a}}
\end{equation}{}
where $\Omega_{\rm m}(a)$ is the matter density at $a(t)$, and it also depends on the gravity model. We choose to anchor
the linear growth factor at the CMB $D(a_{\rm CMB})a^{-1}_{\rm CMB}=1$.

The same GW events 
used to measure peculiar velocities (or
some other source) can also serve as tracers of mass overdensities.
The overdensity power spectrum in redshift space
for mode $\vec{k}$ also depends on gravity 
through $P_{\delta \delta }\propto (bD + fD\mu^2)^2$ where $b$ is the GW host galaxies' bias and $\mu\equiv \cos{(\hat{k} \cdot \hat{r})}$ where $\hat{r}$ is the direction of
the line of sight \citep{1987MNRAS.227....1K}. 
The bias is a nuisance parameter in our analysis, so we marginalize over it when inferring $\gamma$. Future analyses of GW events could establish a prior for this parameter, which is likely to depend on the formation channels for the class of mergers considered. The other power spectrum to consider is the galaxy-velocity cross-correlation, which goes as $P_{v\delta} \propto  (bD + fD\mu^2)fD\mu$. For a derivation of the power spectra see \citet{Song_2009}.

\begin{figure*}
\centering
\includegraphics[width=0.7\linewidth]{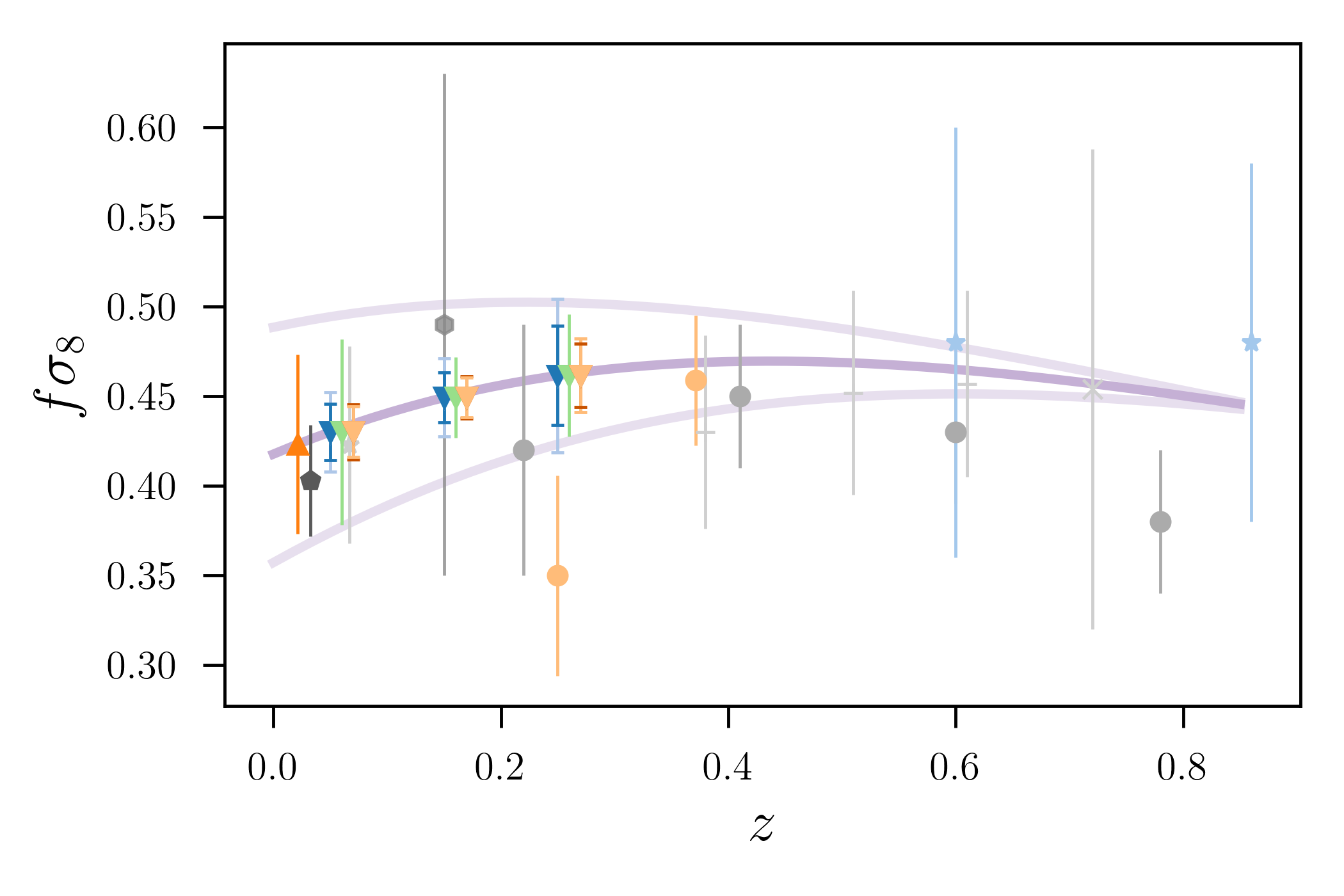}\includegraphics[width=0.3\linewidth]{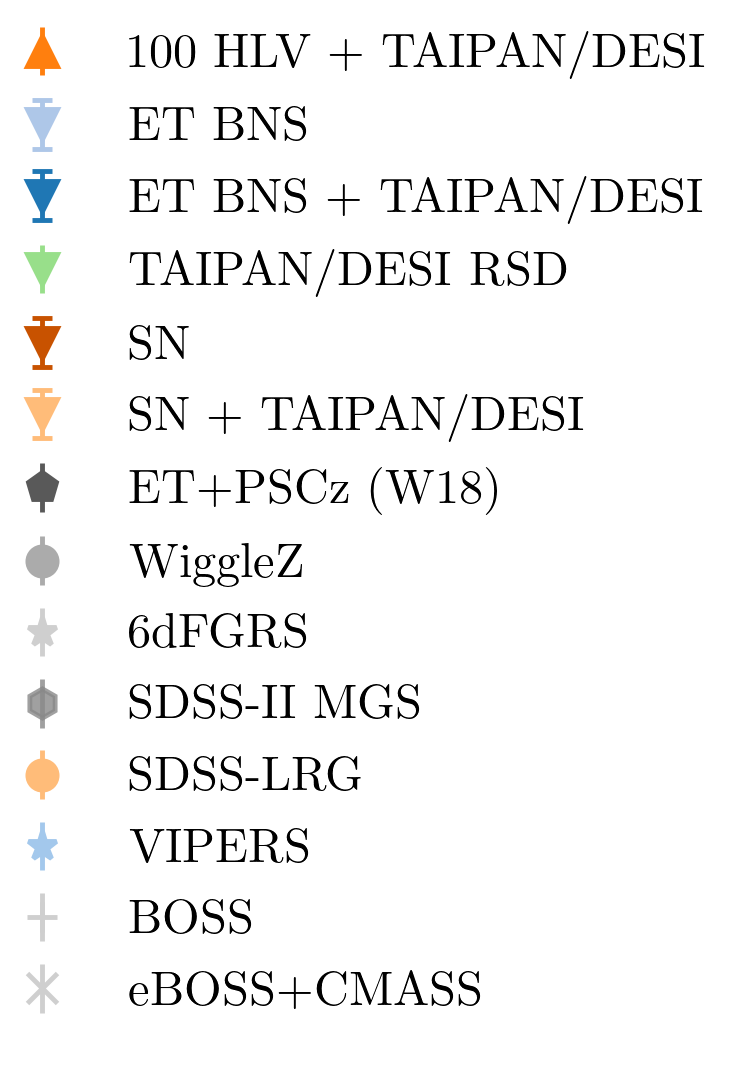}
\caption{Expected constraints on $f(z)\sigma_8(z)$ using peculiar velocities measured using binary neutron star mergers from a 10 year 3G GW experiment (light blue errorbars with triangles) and in combination with DESI/TAIPAN (dark blue errorbars). DESI/TAIPAN RSD--only conservative expected constraints are shown by the green triangles, shifted to higher $z$ for visualization purposes. These are compared to constraints from a 10 year SN survey covering $2\pi$ of the sky that is able to recover a SN magnitude dispersion of $\sigma_M=0.08$ in the same redshift bins (red triangles, shifted to higher $z$ for visualization purposes). The results from a combination of such survey with DESI/TAIPAN RSDs is shown by the light orange triangles, and they are very similar to the SN--only case. Prospects for 100 BNSs from LIGO/Virgo (HLV) at design sensitivity are also reported (darker orange triangle), in combination with galaxy surveys. All the results represented by triangles are computed in this work.
The remaining data points represent existing $f(z)\sigma_8(z)$ measurements from 6dF \citep{6df}, WiggleZ \citep{wigglez}, SDSS-II LRG \citep{sdsslrg}, SDSS-II Main Galaxy sample \citep{sdssmgs}, BOSS \citep{boss}, VIPERS \citep{vipers} and eBOSS-CMASS \citep{eboss}. We also report the forecast from \citet{2018PhRvD..98f3503W} (black pentagon), who use a combination of GW and PSC$z$ galaxy survey data. The darker purple line is the theoretical prediction for $f(z)\sigma_8(z)$ in a Flat $\Lambda$CDM Universe with $\gamma=0.55$ (GR), while the other curves show the theoretical prediction from $\gamma=0.42$ and $\gamma=0.68$ (which are the values predicted for $f(R)$ and DGP gravity, respectively).}\label{fig:fs}
\end{figure*}

Forecasts on the growth index are then computed using the Fisher matrix formalism, following \citet{Howlett_2017a} and \citet{Howlett_2017}. The Fisher information matrix can be written as:
\begin{align}
F_{ij} 
& = \frac{\Omega}{8\pi^2} \int_{r_{\rm min}}^{r_{\rm max}}  \int_{k_{\rm min}}^{k_{\rm max}}  \int_{-1}^{1} r^2 k^2 \text{Tr}\left[ C^{-1} \frac{\partial C}{\partial \lambda_i} C^{-1}
\frac{\partial C}{\partial \lambda_j} \right] d\mu\,dk\,dr
\label{fisher:eqn}
\end{align}
where
\begin{equation}
C(k,\mu,a)  =
  \begin{bmatrix}
   P_{\delta \delta}(k,\mu,a) + \frac{1}{n} &
   P_{v\delta}(k,\mu,a)  \\
   P_{v\delta}(k,\mu,a)  &
  P_{vv}(k,\mu,a) + \frac{\sigma_{v,\text{eff}}^2}{n}
   \end{bmatrix},
\label{cov:eq}
\end{equation}
$\Omega$ is the solid angle over which GW sources are detected, $n$ is the GW events number density, $r$ is the comoving distance and $r_{\text{max}}$ (corresponding to the redshift $z_{\text{max}}$) is the maximum comoving distance at which events are detected.
The minimum distance,
$r_{\text{min}}$, is set to $z=0.01$ for SNe
to ensure small propagated velocity uncertainties, and to $z=0.001$ for GWs because the distance uncertainty is smaller than SNe, especially at these low distances. However, we find that our results are not significantly affected by a more conservative $z=0.01$ cut.

One parameter set we consider
normalizes growth to
the amplitude of clustering today, such
that
$D(z) = \sigma_8(z)$, noting
that the shape of the power spectrum
does not change in our model.
In this case, $\lambda \in \{\{\langle f\sigma_8\rangle\},  b\sigma_8\}$
where 
$\langle f\sigma_8\rangle$'s refer to the effective $f\sigma_8$'s in a set of redshift bins,
and the combination
$b\sigma_8$ is constant in all redshift bins.
We also consider the parameter set $\lambda \in \{\gamma, \Omega_{\rm m0}, b\}$
where $\Omega_{\rm m0}$ is the matter density at $z=0$. Taking $\Lambda$CDM as our fiducial model, 
$\Omega_{\rm m}=\frac{\Omega_{\rm m0}}{\Omega_{m0} + (1-\Omega_{\rm m0})a^3}$.  
The uncertainty in the growth index is then given by $\sqrt{F^{-1}_{\gamma \gamma}}$.

{The 
$\sigma_{v,\text{eff}}^2$ in the shot-noise term is the
velocity variance derived from
the ensemble of sources within a
differential cell volume \citep{2008MNRAS.389.1739A,2017MNRAS.471..839A}.  For indicators whose distance
uncertainties are dominated by intrinsic
magnitude dispersion, all objects in
the same cell share the same relative
distance uncertainty and $\sigma_{v,\text{eff}}^2 = \sigma_v^2$
(see Eq.~\ref{eq:sigma}).
In the case of GW sources, whose
different inclination angles result in a broad range
of distance uncertainties, the 
appropriate number to use for the effective
per-object velocity variance is
$\sigma_{v,\text{eff}}^2= \langle \sigma_v^{-2} \rangle^{-1}$.}

For the Fisher calculation
we take $\gamma=0.55$, $\Omega_{m0}=0.3$, $b=1.2$ (as expected from \citealt{taipan}),
and the equivalent $f\sigma_8$.
We use an external prior
for  $\Omega_{m0}$
with 0.005 uncertainty.
The matter power spectrum
at $P_{\delta \delta}$ at $z=0$ is calculated using the default configuration of
CAMB \citep{Lewis:2002ah}. This is then propagated to redshift $z$ through $D(z;\gamma)$, as explained in detail in Appendix \ref{appendix:sec}.

We take that local
non-linear flows contribute a random
uncorrelated velocity dispersion of
300 km\,s$^{-1}$. The per-object peculiar velocity uncertainty $\sigma_v$ is then related to the distance uncertainty through:
\begin{equation}
    \sigma_v^2 = \Big(1-\frac{1}{aH\chi}\Big)^{-2}   \Big( \frac{\sigma_d}{d} \Big)^2 +
    \left(\frac{300\ \text{km}\,\text{s}^{-1}}{c} \right)^2,\label{eq:sigma}
\end{equation}
where $a$ is the scale factor, $H$ is the Hubble parameter and $\chi$ is the comoving distance, all computed at the distance $d$.

In the following, we will assume that the fractional relative distance uncertainty from the GW experiment scales as the inverse of the SNR  of the GW signal $\rho$:
$\sigma_d/d \propto 1/\rho \propto d$ (e.g. \citealt{Cai_2017,2018arXiv181111723M}), 
and show results for a range of different normalizations to this relation. We ignore the weak-lensing contribution to the distance uncertainty since it does not have a significant contribution at the redshifts considered here. 

We also show results for specific network configurations that have been studied in the literature. At the time of writing, realistic distance uncertainty
distributions for 3G experiments such as those presented in \citet{zhao} were not available. We therefore use approximations for $\sigma_{v,\text{eff}}^2/n$ in Eq.~\ref{cov:eq}, based on the available information. 

The first approximation is to use only the mean $\langle (\sigma_d/d)^{-2}\rangle^{-1}$ in redshift bins to describe a sample of events with a distribution of different $(\sigma_d/d)$:
\begin{align}
\sigma_{v,\text{eff}}^2 &= \langle \sigma_v^{-2} \rangle^{-1}\\
& \approx \Big(\frac{1}{aH\chi}\Big)^{-2} \langle    \Big( \frac{\sigma_d}{d} \Big)^{-2}\rangle^{-1} 
\end{align}
This approximation is useful when considering the ET configuration (three interferometers with $60$ deg opening angles and 10 km arms, arranged in a triangle) studied in \citet{Zhao_2011}. We use their fit for the redshift dependence of $ \langle  ( {\sigma_d/d} )^{-2}\rangle^{-1}\equiv A(z)^{-1}$ (their Eq. 32), which holds for a subsample $n= (1-\cos{20^{\rm o}})~n_T$ of face--on systems with inclination angles $\iota <20$ deg, out of the total number of systems $n_T$. This is a pessimistic scenario
as the low-inclination subset has
the worst distance precision \citep{zhao,Chen_2019}, and
the full population must have
a smaller $\sigma_{v,\text{eff}}^2$.
As a less pessimistic bound, we use
the above approximation for $\sigma_{v,\text{eff}}^2$, but
we use the full population $n_T$.

The second estimate that we consider consists in limits. From information such as the median, we know that a
subpopulation $f$ e.g. 0.5 of the GW-population
distances would have $ \frac{\sigma_d}{d}< 0.01 $
at $z=0.1$.
Then we can
estimate $$\frac{\sigma_{v,\text{eff}}^2}{n}
\lesssim \frac{
 \Big(1-\frac{1}{aH\chi}\Big)^{-2}   \Big( 0.01^2 +
    \left(\frac{300\ \text{km}\,\text{s}^{-1}}{c} \right)^2\Big)}{fn_T}.$$

Note that this is a pessimistic
estimate when considering the median, given that the distribution
does go to lower uncertainties for edge--on
GW systems, and that we only use a fraction of the available data. For this reasons we treat it as an upper limit.

\section{Results}\label{sec:results}

\subsection{Growth of structure in GR}

We first  compute the expected constraint on $f\sigma_8$ in three redshift bins in the range $0.0<z<0.3$.
The fiducial $f\sigma_8$
values correspond to the
expectation from $\Lambda$CDM.
 
Our result for a 10 year ET configuration as in \citet{Zhao_2011} is $\sigma(f\sigma_8)/f\sigma_8=0.0513,~ 0.0485,~ 0.0921$ at $z=0.05,~0.15$ and 0.25, respectively, as shown in Figure \ref{fig:fs} by the blue triangles with light blue errorbars. The best constraint is reached in the second bin because at the lowest redshifts cosmic variance and intrinsic velocity
dispersion give a larger contribution to the final uncertainty, while in the highest redshift bin the relative distance uncertainty is $\sim 3$ times larger than in the second bin. This result is compared to our expected constraints from a 10 year SN survey covering $2\pi$ of the sky that is able to recover a SN magnitude dispersion of $\sigma_M=0.08$ (red triangles). 
Although GW distances
can be significantly better than
those of SNe at $z<0.1$, the SN survey shows slightly better constraints over the first redshift bins because of the higher rate of these transients. The difference between the two results becomes more significant in the last bin, because the SN precision on the distance is assumed to be constant with redshift.

In
Figure~\ref{fig:fs} these constraints (light blue triangles, outer error bars)
are compared
 to existing $f(z)\sigma_8(z)$ measurements from 6dF \citep{6df}, WiggleZ \citep{wigglez}, SDSS-II LRG \citep{sdsslrg}, SDSS-II Main Galaxy sample \citep{sdssmgs}, BOSS \citep{boss}, VIPERS \citep{vipers} and eBOSS-CMASS \citep{eboss}.

The above results show that GW distances and hosts \emph{alone} can already place interesting constraints on  $f\sigma_8$. More precise results can be reached by the addition of a dense galaxy survey, such as those mentioned in Section \ref{sec:experiments}.
The number density of events $n$ considered in the overdensity power--spectrum in Eq.\ (\ref{cov:eq}) is replaced by the much larger galaxy number density $n_g$.
Our Fisher matrix constraints are reported in Table \ref{tab:results} and shown by the dark blue errorbars in the Figure. With a $\sim3\%$ precision at $z\lesssim 0.2$, these bounds are competitive with the constraints expected from the aforementioned SN experiment, and with the forecast by \citet{Howlett_2017a} for a combination of TAIPAN with HI surveys WALLABY (Widefield ASKAP L-band Legacy All-sky Blind Survey; \citealt{2008ExA....22..151J}) and Westerbork Northern Sky HI Survey (WNSHS). In Table \ref{tab:results}, we also report the expected constraints from SNe and RSD from a TAIPAN/DESI--like experiment. The addition of galaxies brings marginal improvement in the first two $z$ bins, while the third bin is better constrained by the SN hosts' rather than the survey galaxies' overdensity power spectrum. This is due to the fact that for a 10 year survey, the number density of SNe is $\sim 5$ times larger than $n_g=10^{-4}~h^{3}~{\rm Mpc}^{-3}$.

The green triangles show our expected constraints from RSD only for a galaxy survey with the number densities mentioned above. It is clear that the GW measurement can bring significant additional information to RSDs in the lowest redshift bin (reducing the RSD--only $1\sigma$ by $\sim 70\%$), while the two probes provide similar constraints in the remaining bins, where their combination provides a 38 and $20\%$ improvement respectively.
 Note that these results hold for an ET---only configuration. If a larger network of 3G detectors is built, and can reach an average 1\% uncertainty in distance at $z=0.1$ (as it is realistic for three detectors), the same constraints can be reached after only $\sim 5$ years.

\begin{table*}[]
\centering
\begin{tabular}{cccc}
Data & $\frac{\sigma(f\sigma_8)}{f\sigma_8}(z=0.05)$  & $\frac{\sigma(f\sigma_8)}{f\sigma_8}(z=0.15)$ & $\frac{\sigma(f\sigma_8)}{f\sigma_8}(z=0.25)$ \\
\hline
ET GW BNS & 0.0513 & 0.0485 & 0.0921 \\ 
ET GW BNS + TAIPAN/DESI & 0.0365 & 0.0311 & 0.0598 \\ 
SN  & 0.0357 & 0.0261 & 0.0381 \\ 
SN + TAIPAN/DESI & 0.0327 & 0.0248 & 0.0445 \\
\end{tabular}
\caption{Expected constraints on $f(z)\sigma_8(z)$ using peculiar velocities measured using binary neutron star mergers from a 10 year ET and in combination with DESI/TAIPAN out to $z< 0.3$. For comparison, we also report constraints from a 10 year SN survey.}\label{tab:results}
\end{table*}

GW forecasts have been previously
made by \citet{2018PhRvD..98f3503W} (black pentagon), who use a similar method to what proposed in this work. We note that the number of events they use out to $z=0.045$ corresponds to a $\sim 3-4$ year experiment with the configuration assumed here, but the competitive constraints are achieved by combining the GW data with the Point Source Catalog redshifts (PSC$z$; \citealt{pscz,branchini}) galaxies. 
Unlike this article, they
do not use velocity-velocity correlations to inform $f\sigma_8$.
We consider
a \citet{2018PhRvD..98f3503W}-like
survey in our formalism
by considering the PSC$z$ number density out to $z=0.045$, and a 3 year GW experiment with distance errors $\sigma_d/d=1\%$ for all events out to $z=0.045$. We find that $\sigma(f\sigma_8)/f\sigma_8= 0.0474$
, versus $\sigma(f\sigma_8)/f\sigma_8= 0.0769$ 
 from \citet{2018PhRvD..98f3503W} (38\% improvement).

In addition, we show constraints from 100 events detected by HLV at design sensitivity, combined with a galaxy survey with $n_g=10^{-3} ~h^{3} ~{\rm Mpc}^{-3}$. We find that this constraint is mostly dominated by the overdensity power spectrum, with the peculiar velocities only providing a $\sim 3\%$ improvement. When additional constraints on the viewing angle are available, the improvement can reach 8--15\% if the the constraint is $5-10$ deg at $1 \sigma$, or if the viewing angle has an upper bound of 30 deg or less, using the derived distance precision from \citet{chen17}.

The dark purple line in Figure \ref{fig:fs} is the $f(z)\sigma_8(z)$ evolution computed from theory assuming a Flat $\Lambda$CDM with $\gamma=0.55$ (GR), while the other curves show the theoretical prediction from $f(R)$ and DGP gravity. It is clear that the measurement proposed in this work will allow us to place interesting constraints on these gravity models.

\subsection{Testing GR}

\begin{figure*}
\centering
\includegraphics[width=0.48\linewidth]{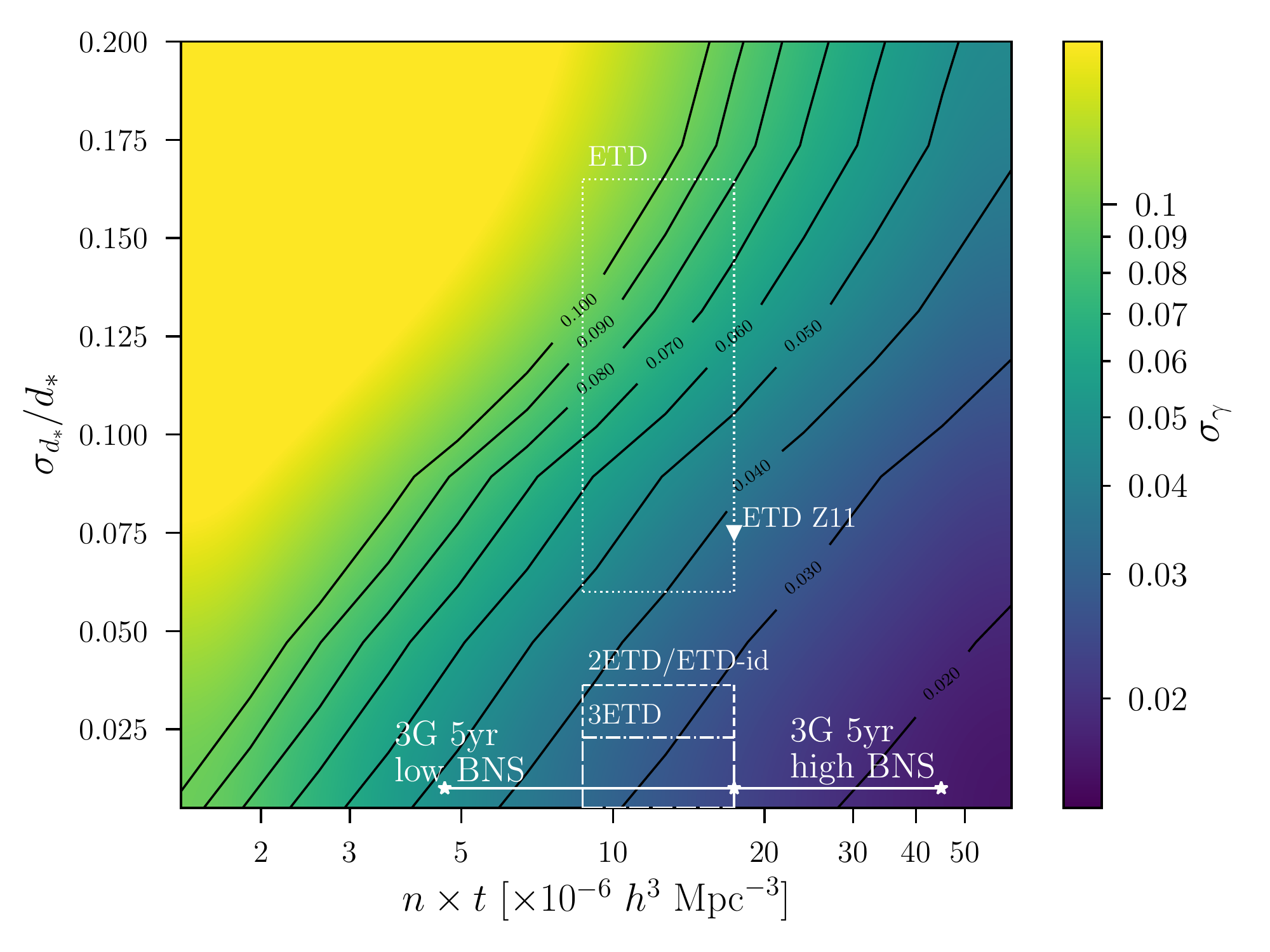}
\includegraphics[width=0.48\linewidth]{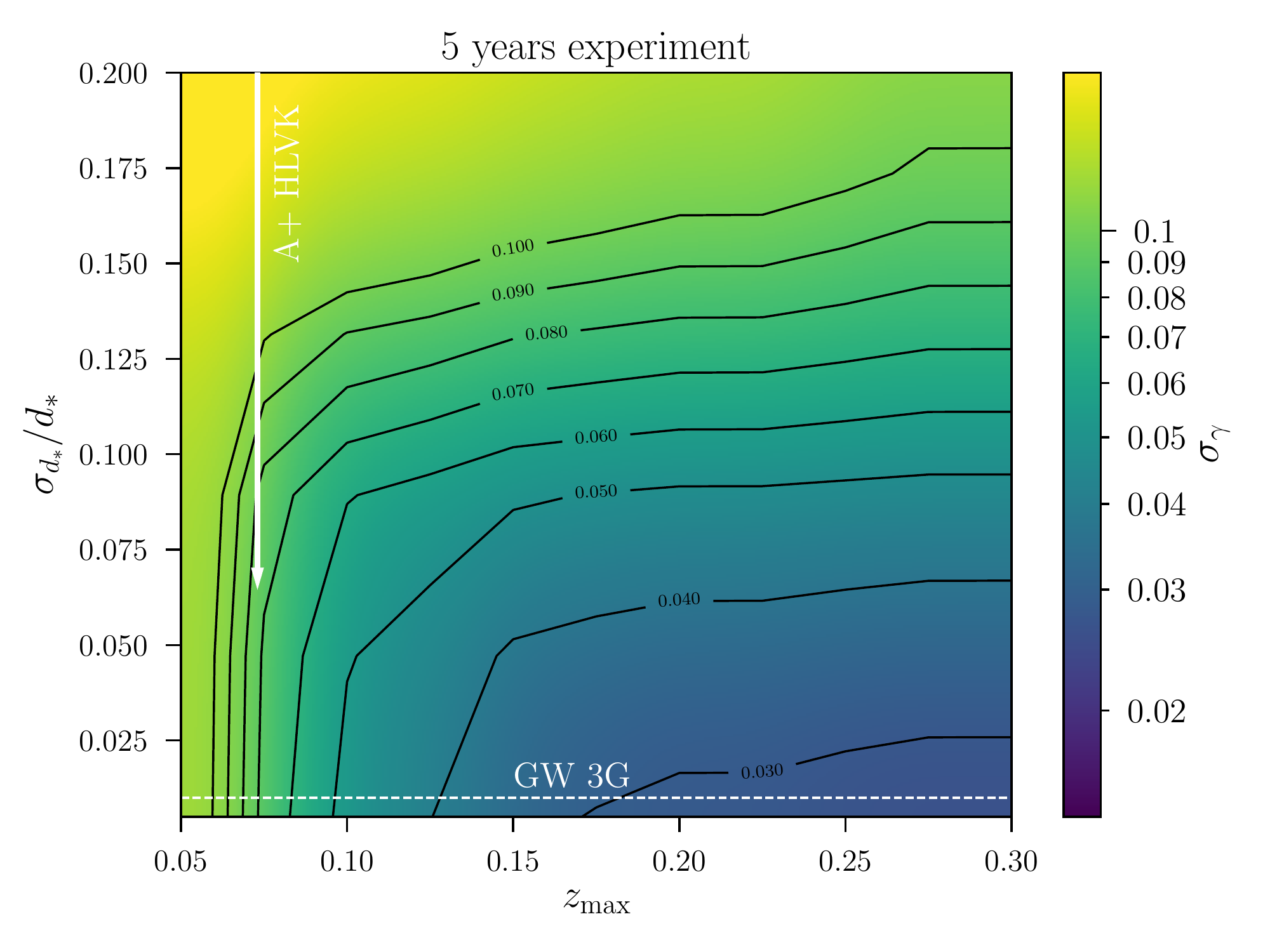}
\caption{Growth index uncertainty for different values of distance precision as a function of BNS volumetric rates integrated over time out to $z_{\rm max}<0.3$ (\emph{left panel}), and as a function of the maximum redshift $z_{\rm max}$ out of which we consider GW events. The distance uncertainty $\sigma_{d_*}$ is the uncertainty at a reference distance $d_*$, here corresponding to $z_*=0.1$. The BNS volumetric rate considered for the specific estimates on the left, and all points on the right panel, is the maximum \emph{a posteriori} value from the latest GW estimate, $1.09\times 10^{-6}\times(h/0.679)^3$ Mpc$^{-3}$ yr$^{-1}$ \citep{190425}. On the left panel, also the low and high $90\%$ CL limits in the rate are shown for an ideal 3G experiment (where the distance precision is 1\% at $z=0.1$) after 5 years. The effect of the shift in number density from the high to the low bound of the 90\% CL range is shown by the white line. The boxes represent limits within which we expect the constraint to fall for the various ET configurations studied in \citet{zhao}. The white triangle shows our result using the \citet{Zhao_2011} approximation for 1 ET. The distance precision for the ideal 3G experiment is also shown for reference on the right panel by the dashed line. The arrow shows possible constraints from the A+ LIGO/Virgo/KAGRA network, if improved viewing angle constraints can be derived from EM observations, reducing the distance uncertainty from order $\sim 20\%$ down to $10\%$ or better. }\label{fig:surfaces}
\end{figure*}

In the second part of this work, in
place of determining $f\sigma_8$
in different redshift bins,
we let $\gamma$ be a free parameter. Expected constraints on the growth index are shown in Figure \ref{fig:surfaces}. The left hand plot shows the precision for different values of
effective fractional distance uncertainty for different volumetric rates integrated over time (i.e. $n \times t$). The distance uncertainty $\sigma_{d_*}$ is the uncertainty at a reference distance $d_*$, here corresponding to $z_*=0.1$. Events at distances different from $d_*$ have a distance precision that scales as described in Section \ref{sec:experiments}. The number of events considered corresponds to the total of an experiment length up to 18 years for a BNS volumetric
restframe rate corresponding to the maximum \emph{a posteriori}
from the latest estimate, $1.09\times 10^{-6}$ Mpc$^{-3}$ yr$^{-1}$ \citep{190425}.

If the rate is lower, it will take more time to reach the same constraints. The effect of the shift in number density from the high to the low bound of the 90\% CL range is shown by the white line on the left hand plot for a 5 years experiment, for GW sources out to $z=0.3$, with 
a distance precision of 1\% (or following a distribution of distance uncertainties with $\langle
(\sigma_d/d)^{-2} \rangle^{-1/2}=0.01$)
at $z=0.1$. 
We find that data from a 5 year GW experiment that is able to reach a $\lesssim$ few per cent precision on average for sources at $z=0.1$ will be enough to constrain the growth index to $\sigma(\gamma)\lesssim 0.04$ if the BNS rate is on the high end, allowing us to discern between GR and other popular gravity models. This precision level is in fact reached by the ET configuration studied in \citet{Zhao_2011}. When using their approximation for the redshift dependance of the distance uncertainty, we recover $\sigma(\gamma)= 0.039$ after 5 years if we detect EM counterparts out to $z_{\rm max}=0.3$, and $\sigma(\gamma)= 0.041$ if $z_{\rm max}=0.2$.

Expected limits for other different configurations of detectors (estimated with the approximation described in Section \ref{sec:method} from the median of a population) are given by the boxes in Figure \ref{fig:surfaces}. Using the median values from \citet{zhao}, we can provide an extreme pessimistic bound using only the best $50\%$ of events,
that will have a distance precision equal or better than what reported on the $y-$axis. The extreme optimistic bound is chosen as the lowest precision considered in \citet{zhao} for the redshift bins in consideration. The other two bounds of the boxes can be drawn by considering all sources having the median and lowest precision of the sample, respectively.
The configurations considered include one ET detector, one ET with ideal low--frequency noise described in \citet{zhao}, 2 ET detectors (which is similar to the ET+CE case), and the 3 ET (or 3CE) network.

 The right hand side plot of Figure \ref{fig:surfaces} shows the expected precision on $\gamma$ for different values of distance uncertainty and different maximum redshift values, starting from the same assumptions of the left hand side plot. The rapid flattening of the $\sigma_\gamma$ contours with the redshift show that, for a given distance precision, adding events at higher redshifts only very slowly helps constraining the growth index, due to the increasing distance uncertainty for more distant events.  The assumption that we have made on the detectability of KNe out to $z\sim 0.3$ can be therefore relaxed to reach only $z\sim 0.15$ for an experiment with $\sigma_{d_*}/d_*= 5\% $ or $z\sim 0.2$ for an experiment with $\sigma_{d_*}/d_*= 1\% $, since the constraining power flattens out at higher redshifts for these distance precisions, and in both cases recovering $\sigma(\gamma) \lesssim 0.04$.
On the low-redshift
 end of the Figure, for
 $z_{\rm max }\lesssim 0.05$  BNSs cannot even reach a 0.1 uncertainty on $\gamma$ and thus cannot provide an interesting constraint using LIGO/Virgo at design sensitivity. The A+ upgrade of current detectors could provide more interesting constraints at the level of $\sigma(\gamma) \gtrsim 0.07$ only if improved distance measurements can be achieved by EM constraints on the geometry of the systems, as mentioned in Section \ref{sec:experiments}. This can be seen by following the arrow in the right panel of Figure \ref{fig:surfaces}. However, such a measurement would not be precise enough to discern between popular gravity models. Even if the
 measured distance precision improves,
 at low-redshift the 300 km\,s$^{-1}$
 random noise and sample variance
 dominate the dispersion in peculiar
 velocities.
 
 Similarly to what presented in the previous subsection, we combine the peculiar velocity field measured from GWs with the overdensities from a RSD galaxy survey.  In this case, the constraints are improved to $\sigma_\gamma\sim 0.02-0.03$ when combined with galaxy overdensities.
On the other hand, a combination of RSD surveys covering $\sim 90\%$ of the sky such as DESI+TAIPAN can only reach $\sigma_\gamma\sim 0.05$ at $z<0.3$. However, note that DESI is expected to constrain $\sigma_\gamma=0.026$ by measuring the growth of structure at larger redshifts out to $z<1.6$ \citep{kim2020}.

\begin{figure}
\centering
\includegraphics[width=1\linewidth]{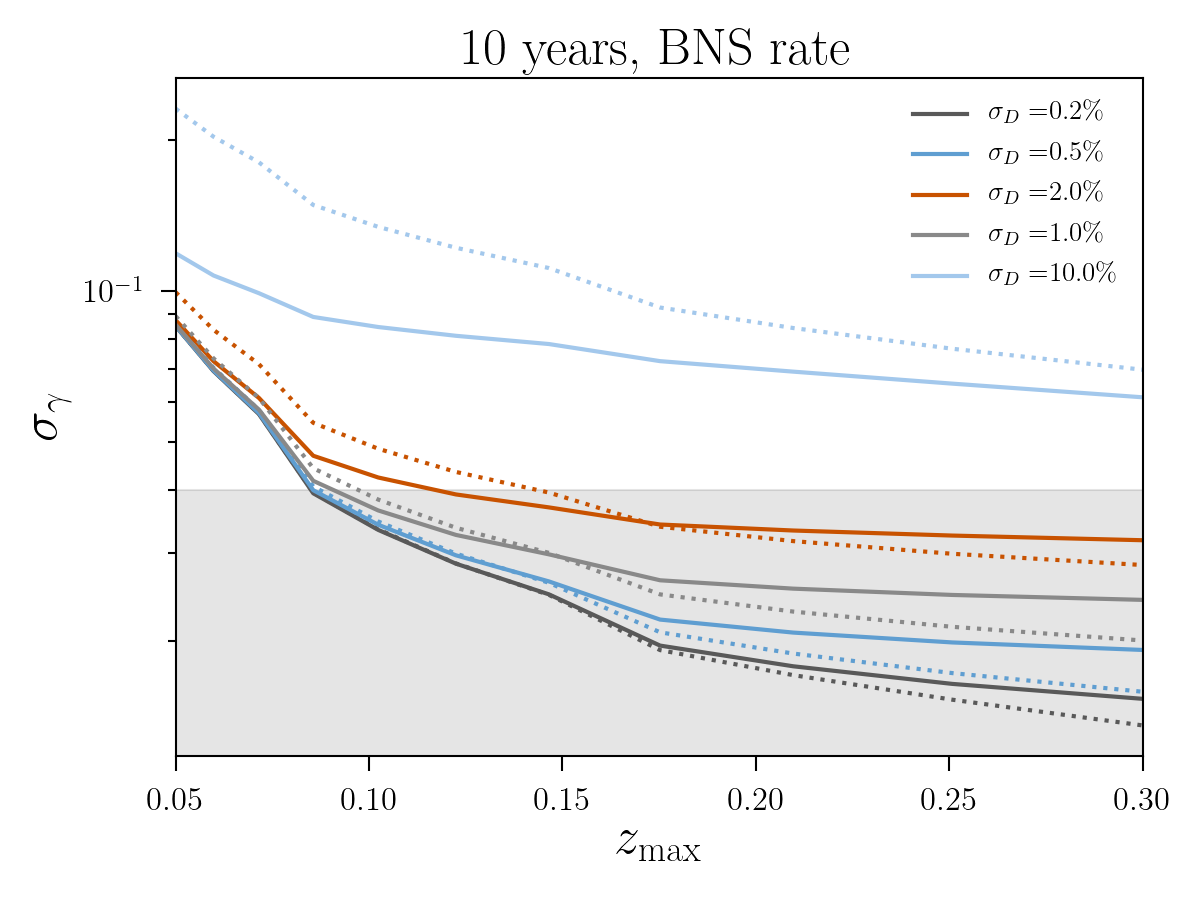}
\caption{Comparison between a SN (dashed lines) and a GW (solid lines) experiment to measure the growth index from peculiar velocities, for different distance uncertainties at $z=0.1$, as a function of the maximum distance reach of the SNe/GW mergers.  In this idealized case, BNSs and SNe have the same value of $\frac{\sigma_v^2}{n}$ at $z=0.1$, implying that the reported $\sigma_d/d$ precision corresponds to a magnitude uncertainty at all redshifts $\sigma_M=\frac{5}{\ln{(10)}}\frac{\sigma_d}{d}$ for the SNe. Both SN and GW experiments cover $90\%$ of the sky.
 Differences in the $\gamma$ constraints are driven by the different redshift dependence of the distance uncertainty measurement. The shaded region shows where constraints on $\gamma$ can discern between popular gravity models ($\sigma_\gamma \lesssim 0.04$).}\label{fig:comparison}
\end{figure}

In Figure \ref{fig:comparison} we show a comparison between the constraints from a GW and a SN experiment for different distance uncertainties. In this idealized case, BNSs and SNe experiments cover $90\%$ of the sky for 10 years, and they have the same $\frac{\sigma_v^2}{n}$ entering in Eq.(\ref{cov:eq}) at $z=0.1$. In other words, the reported $\sigma_d/d$ precision corresponds to a magnitude uncertainty at all redshifts $\sigma_M=\frac{5}{\ln{(10)}}\frac{\sigma_d}{d}$ for the SNe. The different behavior observed between the GW and the SN approach in measuring the growth index is driven by the different scaling of the distance measurement precision with redshift. The redshift dependence of the distance uncertainty makes the GW measurements more precise at $z<0.1$, thus experiment is more constraining in $\gamma$ than the SN case at the lowest redshift. At higher redshifts, adding more events does not improve significantly the constraints from GWs because the distance uncertainty is much less constraining than the events around $z\sim 0.1$, and the curve starts to flatten. This is not the case for SNe, since the intrinsic scatter in the magnitude does not depend on redshift, so including higher $z$ events still provides a significant contribution.
We also note that at $z<0.1$ the expected constraint on $\gamma$ is very similar for all distance uncertainties $\sigma_d<1\%$, and for both the SN and GW cases. At low-redshifts $z<0.05$, the advantage of a small $\sigma_d/d$
quickly saturates out due to sample variance.

\section{Conclusions}\label{sec:conclusion}

In this article, we presented an alternative method to test GR and constrain the growth of structure by measuring the galaxies' peculiar velocity field using gravitational wave measurements of compact binary mergers. We find that data from a 5 year GW experiment, in conjunction with follow--up facilities, that are able to detect BNS mergers out to $z\sim0.2$ and reach a $\lesssim$ few per cent precision on average for sources at $z=0.1$ will constrain $f\sigma_8$ to $\sigma(f\sigma_8)\sim 3-4\%$ in that redshift range. For a single ET, $\sigma(f\sigma_8)\sim 5\%$ is reached in 10 years.

Moreover, this method will constrain the growth index to $\sigma(\gamma)\lesssim 0.04$, allowing us to discern between GR and other popular gravity models at $\sim 3 \sigma$. 
The distance precision and the detection horizon fit well with what is expected for a network of 3G GW detectors, such as ET and CE. On the other hand, the potential of the cosmological probe presented here highlights the need for extensive optical to near-infrared follow-up of nearby BNSs to identify the associated kilonovae. 
Once identified, it is likely that large number of host galaxies for these events will be already observed by planned spectroscopic surveys  (DESI, 4MOST, TAIPAN) in the coming decade. 

These results show that a self-contained 3G GW experiment combined with adequate optical--NIR follow--up efforts can provide interesting results to probe the growth of structure in the local Universe.
Nevertheless, galaxy surveys 
are much more effective at measuring
the overdensity field, and we
find that they are complementary to the peculiar velocity field probed by the GWs. When we combine GWs and DESI+TAIPAN--like surveys, $f\sigma_8$ can be constrained to $\sim 2-3$\% at $z< 0.2$ depending on the network configuration, and $\sigma(\gamma)=0.02-0.03$, which could be a decisive test for General Relativity. We also find that the same RSD surveys alone can only reach $\sigma_\gamma\sim 0.05$ at these redshifts, demonstrating the value of adding peculiar velocities. Note that these results are competitive with other probes from upcoming experiments, such as from a combination of SNe from LSST and DESI \citep{kim2020}.

There is a significant gain in constraining power ($\sim 38\%$ at $z<0.045$) when considering the peculiar velocity power spectrum and the overdensity power spectrum in addition to their cross--correlation. This is an advantage of the method presented here, compared to the one proposed in \citet{2018PhRvD..98f3503W}.

Our results show that events from LIGO/Virgo at design sensitivity cannot provide interesting constraints on the growth of structure, unless improved distance precision from EM observations can be pursued to provide a $\lesssim10\%$ distance precision for $\mathcal{O}(100)$ events, and these are combined with galaxy surveys overdensities. If this is possible, a $\gtrsim 10\%$ improvement on the RSD--only 1$\sigma$ measurements on $f\sigma_8$ from upcoming galaxy surveys can be achieved. Improved constraints on the viewing angle could improve prospects for the proposed method during the current decade. Optical-NIR data could be a particularly powerful tool for this purpose, since kilonovae can be observed from all directions and thus can possibly be identified for most GW detections, unlike counterparts at other wavelengths \citep{2012ApJ...746...48M}.

In this work, we focused on BNS mergers. However, there are additional compact object mergers that could improve the statistics presented. In particular, it has been shown that neutron star--black hole mergers (NSBH) can be accompanied by an EM counterpart (e.g. Kasen 2017), and that they can provide improved distance constraints compared to BNSs \citep{vitale}, leading to improved cosmological constraints. Also NSBH and binary black hole mergers without counterpart can contribute to this analysis if their localization is so accurate to only fit one galaxy (a small fraction of events is expected to satisfy this condition already with current generation detectors; \citealt{chen12}). 

Note that the luminosity distances from GWs discussed in this work are estimated assuming GR. The method presented here aims at testing GR, rather than constraining different gravity models. A significant deviation of $\gamma$ from the value expected in GR using this method, would mean that a modification of gravity exists. On the other hand, a result consistent with $\gamma=0.55$ could still hide modifications of gravity due to the GR assumption in the GW distance estimates. In this case, inconsistencies in $\gamma$ with precision measurements from EM--based distances would hint at such modification, since the luminosity distance in the GW measurement would have a different meaning (e.g. \citealt{Saltas_2014,Nishizawa_2018,2018PhRvD..97j4066B}). However, \citet{linder2020limited} shows that differences between GW and EM distances are at the sub-percent level in the local Universe for popular gravity models, so we expect this effect to be negligible at the redshifts probed here. 

In the future, a simultaneous measurement of the peculiar velocity field and the Hubble diagram using gravitational waves as standard candles, will fully exploit the cosmological information enclosed in the connection between gravitational wave sources and the large scale structure of the Universe. 

\acknowledgments
\input{ack}

\appendix

\section{Contrast between our analysis and that of Wang et al.}\label{app:comparison}

Conservation of mass establishes a
relationship between density and momentum
fields through the continuity equation,
which to first order gives
\begin{align}
        Haf\delta(\mathbf{x}) + \nabla \cdot \mathbf{v}(\mathbf{x})& =0 \nonumber\\
            Ha\beta\delta^g(\mathbf{x}) + \nabla \cdot \mathbf{v}(\mathbf{x})& =0.
            \label{continuity:eqn}
\end{align}
\citet{2018PhRvD..98f3503W} use this relationship
between galaxy overdensity and velocity fields to
determine $\beta=f/b$.  Combining this with $\sigma_{8,g}$ as measured from the galaxy survey yields $f \sigma_8 = \beta \sigma_{8,g}$.
Conceptually (though not exactly in practice), 
the model for the data includes $\beta$
and a model for the underlying galaxy overdensity
field.  Eq.~\ref{continuity:eqn} is used
to calculate the velocity field (taking
care with the constant of integration). 
While \citet{2018PhRvD..98f3503W}
do not assign any error to this process,
the similar analysis of
\citet{2019arXiv191209383B} quadratically adds
a $150$~km\,s$^{-1}$ uncorrelated uncertainty to each velocity measurement.

The approach in this article uses the
above comparison of density
and velocity fields and adds independent information from the anisotropy power
spectrum at the time of the CMB.
A local peculiar velocity survey does not
cover the same volume of the CMB so a direct
comparison of the density and velocity fields
is not possible.  Nevertheless, the correlations in
matter underdensities in the different
volumes
should be consistent.

The diagonal terms of the matrix
in Equation~\ref{cov:eq} contain
the autocorrelations in
galaxy counts and peculiar velocities for a single $k$-mode
measured in Fourier space as
predicted from the anisotropy power
spectrum at the time of the CMB.  We include in the velocity shot-noise a 300~km\,s$^{-1}$
term that represents sources
of peculiar velocity that are not represented in our model e.g.\ contributions
from  $k$-modes excluded
in our calculations.  It does not include errors
that are in common between galaxy
counts and velocities, for example
differences between CAMB predictions
and the true overdensity field today.
The off-diagonal term
gives the cross-correlation
that accounts for the fact that galaxy counts and velocities
arise from the same overdensity
field.  We do not add an error that
corresponds to
\citet{2019arXiv191209383B},
but by limiting our
calculations to $k_{\text{min}} \le k \le k_{\text{max}}$ we use less
of the available data relative to their analysis,
which uses all the information
in real space.

\section{Relationships Between the Peculiar Velocity
and Peculiar Magnitude Power Spectra
and the CMB Matter Power Spectrum}
\label{appendix:sec}
Predictions for the overdensity
power spectrum are made by solving
the Boltzmann Equations.
For our Fisher calculations, we do not
recalculate this power spectrum as our model
parameters change.
While imperfect, this approach is
motivated as follows.
In linear theory  the linear growth factor $D$ is introduced
to represent the time evolution of the density field.  Its
normalization is arbitrary, but a common practice is
to normalize it to the CMB, such that $D(z_{CMB})/a_{CMB} =1$.
This choice is useful because it anchors $D$,
independent of the parameters upon which
it may depend (e.g.\ $\Omega_{M0}$, $\gamma$), to a redshift
with a precise measurement of $P_{\delta \delta}$.
With this convention
\begin{equation}
    P_{\delta \delta}(k; z, \gamma) = a^{-2}_{CMB} D^2(z;\gamma) P_{\delta \delta}(k; z=CMB)
\end{equation}
and
\begin{align}
    P_{vv}(k,\mu; z,\gamma)& = \left(\mu Haf \right)^2 k^{-2}  P_{\delta \delta}(k; z, \gamma)\\
    & = \left(\mu HfD \right)^2 \left(\frac{a}{a_{CMB}}\right)^2 k^{-2}  P_{\delta \delta}(k; z=CMB).
\end{align}
For distance indicators it is useful to work with peculiar
magnitudes \citep{2006PhRvD..73l3526H},
i.e.\ magnitude deviations from the background cosmological expansion.  At low redshift and
for small peculiar velocities
the peculiar magnitude power spectrum is
\begin{equation}
    P_{\delta m \delta m}(k,\mu; z,\gamma) \approx \left(\frac{5}{\ln{10}} \right)^2 \left(\mu f \right)^2 \left(d_L k\right)^{-2}  P_{\delta \delta}(k; z, \gamma).
\end{equation}

Going beyond linear theory, CAMB provides a more precise calculation of $P_{\delta \delta}$ using the CMB as an initial
condition and standard gravity.
Using the CAMB power spectrum as reference
\begin{equation}
    P_{\delta m \delta m}(k,\mu; z,\gamma) = \left(\frac{5}{\ln{10}} \right)^2 \left(\mu f \right)^2 \left(d_L k\right)^{-2} 
    \frac{P_{\delta \delta}(k; z, \gamma)}{P^{\text{CAMB}}_{\delta \delta}(k; z=0)}
    P^{\text{CAMB}}_{\delta \delta}(k; z=0).
\end{equation}

For the normalization term, taking $P^{\text{CAMB}}_{\delta \delta}(k; z) \approx P_{\delta \delta}(k; z, \gamma=0.55)
$ and $f=\Omega(z)^\gamma$
\begin{align}
  \frac{P_{\delta \delta}(k; z, \gamma)}{P^{\text{CAMB}}_{\delta \delta}(k; z=0)} & =    \left( \frac{D(z;\gamma)}{D^\text{CAMB}(z=0)} \right)^2 \\
  & \approx    \left( \frac{D(z;\gamma)}{D(z=0;\gamma=0.55)} \right)^2 \\
  & =\exp{
  \left[
  2 \left(\int^a_{a_{CMB}} \Omega_m^{\gamma} d\ln{a} - \int^1_{a_{CMB}} \Omega_m^{0.55} d\ln{a}\right)
  \right]
  }.
\end{align}
For the special case of calculating the 
Fisher Matrix at $\gamma=0.55$ the normalization
of $D$ is irrelevant as
\begin{align}
  \frac{P_{\delta \delta}(k; z, \gamma)}{P^{\text{CAMB}}_{\delta \delta}(k; z=0)} & \approx    \exp{
   \left(2\int^a_1 \Omega_m^{0.55} d\ln{a}\right)
  }.
\end{align}

For some applications, normalizing $D=1$ at $z=0$ is appropriate.  Then
\begin{equation}
    P_{\delta \delta}(k; z, \gamma) =  D^2(z;\gamma) P_{\delta \delta}(k; z=0).
\end{equation}
For this choice of normalization, $P_{\delta \delta}(k; z=CMB, \gamma)$ does depend on $\gamma$ through $D^2(z=CMB;\gamma)$.

\bibliographystyle{yahapj_twoauthor_arxiv_amp}
\bibliography{references}
\end{document}

%% file: ack.tex
We thank Marica Branchesi, Zoheyr Doctor, Jan Harms, Cullan Howlett and Eric Linder for very useful discussion.

Work supported by the Fermi National Accelerator Laboratory, managed and operated by Fermi Research Alliance, LLC under Contract No. DE-AC02-07CH11359 with the U.S. Department of Energy. The U.S. Government retains and the publisher, by accepting the article for publication, acknowledges that the U.S. Government retains a non-exclusive, paid-up, irrevocable, world-wide license to publish or reproduce the published form of this manuscript, or allow others to do so, for U.S. Government purposes.

This work is supported in part by 
the U.S.\ Department of Energy, Office of Science, Office of High Energy Physics, under Award DE-SC-0007867 and contract no.\ DE-AC02-05CH11231.


%% file: main.bbl
\begin{thebibliography}{}
\providecommand\natexlab[1]{#1}
\providecommand\JournalTitle[1]{#1}
\providecommand{\eprint}[1][]{\url{#1}}

\bibitem[{{Abate} {et~al.}(2008){Abate} \& {Bridle} \& {Teodoro} \& {Warren} \&
  {Hendry}}]{2008MNRAS.389.1739A}
{Abate}, A., {Bridle}, S., {Teodoro}, L. F.~A., {Warren}, M.~S., \& {Hendry},
  M. 2008,
  \href{http://dx.doi.org/10.1111/j.1365-2966.2008.13637.x}{\JournalTitle{\mnras},
  389, 1739}, \eprint arXiv:{0802.1935}

\bibitem[{{Abate} \& {Lahav}(2008)}]{abate}
{Abate}, A. \& {Lahav}, O. 2008,
  \href{http://dx.doi.org/10.1111/j.1745-3933.2008.00519.x}{\JournalTitle{\mnras},
  389, L47}, \eprint arXiv:{0805.3160}

\bibitem[{{Abbott} {et~al.}(2017){Abbott} \& {Abbott}
  {et~al.}}]{2017Natur.551...85A}
{Abbott}, B.~P., {Abbott}, R., {Abbott}, T.~D., {et~al.} 2017,
  \href{http://dx.doi.org/10.1038/nature24471}{\JournalTitle{\nat}, 551, 85},
  \eprint arXiv:{1710.05835}

\bibitem[{Abbott {et~al.}(2017)Abbott \& Abbott {et~al.}}]{CE}
Abbott, B.~P., Abbott, R., Abbott, T.~D., {et~al.} 2017,
  \href{http://dx.doi.org/10.1088/1361-6382/aa51f4}{\JournalTitle{Classical and
  Quantum Gravity}, 34, 044001}

\bibitem[{{Abbott} {et~al.}(2018){Abbott} \& {Abbott} {et~al.}}]{LIGOprospects}
{Abbott}, B.~P., {Abbott}, R., {Abbott}, T.~D., {et~al.} 2018,
  \href{http://dx.doi.org/10.1007/s41114-018-0012-9}{\JournalTitle{Living
  Reviews in Relativity}, 21, 3}, \eprint arXiv:{1304.0670}

\bibitem[{{Adams} \& {Blake}(2017)}]{2017MNRAS.471..839A}
{Adams}, C. \& {Blake}, C. 2017,
  \href{http://dx.doi.org/10.1093/mnras/stx1529}{\JournalTitle{\mnras}, 471,
  839}, \eprint arXiv:{1706.05205}

\bibitem[{Barnes \& Kasen(2013)}]{Barnes_2013}
Barnes, J. \& Kasen, D. 2013,
  \href{http://dx.doi.org/10.1088/0004-637x/775/1/18}{\JournalTitle{The
  Astrophysical Journal}, 775, 18}

\bibitem[{{Belgacem} {et~al.}(2018){Belgacem} \& {Dirian} \& {Foffa} \&
  {Maggiore}}]{2018PhRvD..97j4066B}
{Belgacem}, E., {Dirian}, Y., {Foffa}, S., \& {Maggiore}, M. 2018,
  \href{http://dx.doi.org/10.1103/PhysRevD.97.104066}{\JournalTitle{\prd}, 97,
  104066}, \eprint arXiv:{1712.08108}

\bibitem[{{Betoule} {et~al.}(2014){Betoule} \& {Kessler}
  {et~al.}}]{2014A&A...568A..22B}
{Betoule}, M., {Kessler}, R., {Guy}, J., {et~al.} 2014,
  \href{http://dx.doi.org/10.1051/0004-6361/201423413}{\JournalTitle{\aap},
  568, A22}, \eprint arXiv:{1401.4064}

\bibitem[{{Beutler} {et~al.}(2012){Beutler} \& {Blake} {et~al.}}]{6df}
{Beutler}, F., {Blake}, C., {Colless}, M., {et~al.} 2012,
  \href{http://dx.doi.org/10.1111/j.1365-2966.2012.21136.x}{\JournalTitle{\mnras},
  423, 3430}, \eprint arXiv:{1204.4725}

\bibitem[{{Blake} {et~al.}(2011){Blake} \& {Brough} {et~al.}}]{wigglez}
{Blake}, C., {Brough}, S., {Colless}, M., {et~al.} 2011,
  \href{http://dx.doi.org/10.1111/j.1365-2966.2011.18903.x}{\JournalTitle{\mnras},
  415, 2876}, \eprint arXiv:{1104.2948}

\bibitem[{{Boruah} {et~al.}(2019){Boruah} \& {Hudson} \&
  {Lavaux}}]{2019arXiv191209383B}
{Boruah}, S.~S., {Hudson}, M.~J., \& {Lavaux}, G. 2019, \JournalTitle{arXiv
  e-prints}, arXiv:1912.09383, \eprint arXiv:{1912.09383}

\bibitem[{{Branchini} {et~al.}(1999){Branchini} \& {Teodoro}
  {et~al.}}]{branchini}
{Branchini}, E., {Teodoro}, L., {Frenk}, C.~S., {et~al.} 1999,
  \href{http://dx.doi.org/10.1046/j.1365-8711.1999.02514.x}{\JournalTitle{\mnras},
  308, 1}, \eprint arXiv:{astro-ph/9901366}

\bibitem[{Cai \& Yang(2017)}]{Cai_2017}
Cai, R.-G. \& Yang, T. 2017,
  \href{http://dx.doi.org/10.1103/physrevd.95.044024}{\JournalTitle{Physical
  Review D}, 95}

\bibitem[{{Chen} {et~al.}(2018){Chen} \& {Fishbach} \& {Holz}}]{chen17}
{Chen}, H.-Y., {Fishbach}, M., \& {Holz}, D.~E. 2018,
  \href{http://dx.doi.org/10.1038/s41586-018-0606-0}{\JournalTitle{\nat}, 562,
  545}, \eprint arXiv:{1712.06531}

\bibitem[{{Chen} \& {Holz}(2016)}]{chen12}
{Chen}, H.-Y. \& {Holz}, D.~E. 2016, \JournalTitle{arXiv e-prints},
  arXiv:1612.01471, \eprint arXiv:{1612.01471}

\bibitem[{Chen {et~al.}(2019)Chen \& Vitale \& Narayan}]{Chen_2019}
Chen, H.-Y., Vitale, S., \& Narayan, R. 2019,
  \href{http://dx.doi.org/10.1103/physrevx.9.031028}{\JournalTitle{Physical
  Review X}, 9}

\bibitem[{Chornock {et~al.}(2019)Chornock \& Cowperthwaite
  {et~al.}}]{chornock2019multimessenger}
Chornock, R., Cowperthwaite, P.~S., Margutti, R., {et~al.} 2019,
  Multi-Messenger Astronomy with Extremely Large Telescopes, \eprint
  arXiv:{1903.04629}

\bibitem[{da~Cunha {et~al.}(2017)da~Cunha \& Hopkins {et~al.}}]{taipan}
da~Cunha, E., Hopkins, A.~M., Colless, M., {et~al.} 2017,
  \href{http://dx.doi.org/10.1017/pasa.2017.41}{\JournalTitle{Publications of
  the Astronomical Society of Australia}, 34}

\bibitem[{{de la Torre} {et~al.}(2017){de la Torre} \& {Jullo}
  {et~al.}}]{vipers}
{de la Torre}, S., {Jullo}, E., {Giocoli}, C., {et~al.} 2017,
  \href{http://dx.doi.org/10.1051/0004-6361/201630276}{\JournalTitle{\aap},
  608, A44}, \eprint arXiv:{1612.05647}

\bibitem[{{DESI Collaboration} {et~al.}(2016){DESI Collaboration} \&
  {Aghamousa} {et~al.}}]{desi}
{DESI Collaboration}, {Aghamousa}, A., {Aguilar}, J., {et~al.} 2016,
  \JournalTitle{arXiv e-prints}, arXiv:1611.00036, \eprint arXiv:{1611.00036}

\bibitem[{{Dhawan} {et~al.}(2020){Dhawan} \& {Bulla} \& {Goobar} \& {Sagu{\'e}s
  Carracedo} \& {Setzer}}]{dhawan}
{Dhawan}, S., {Bulla}, M., {Goobar}, A., {Sagu{\'e}s Carracedo}, A., \&
  {Setzer}, C.~N. 2020,
  \href{http://dx.doi.org/10.3847/1538-4357/ab5799}{\JournalTitle{\apj}, 888,
  67}, \eprint arXiv:{1909.13810}

\bibitem[{{Dilday} {et~al.}(2010){Dilday} \& {Smith} \& {Bassett} \& {Becker}
  \& {Bender} {et~al.}}]{2010ApJ...713.1026D}
{Dilday}, B., {Smith}, M., {Bassett}, B., {et~al.} 2010,
  \href{http://dx.doi.org/10.1088/0004-637X/713/2/1026}{\JournalTitle{ApJ},
  713, 1026}, \eprint arXiv:{1001.4995}

\bibitem[{Dobie {et~al.}(2020)Dobie \& Kaplan {et~al.}}]{Dobie_2020}
Dobie, D., Kaplan, D.~L., Hotokezaka, K., {et~al.} 2020,
  \href{http://dx.doi.org/10.1093/mnras/staa789}{\JournalTitle{Monthly Notices
  of the Royal Astronomical Society}, 494, 2449–2464}

\bibitem[{{Gordon} {et~al.}(2007){Gordon} \& {Land} \& {Slosar}}]{gordon}
{Gordon}, C., {Land}, K., \& {Slosar}, A. 2007,
  \href{http://dx.doi.org/10.1103/PhysRevLett.99.081301}{\JournalTitle{\prl},
  99, 081301}, \eprint arXiv:{0705.1718}

\bibitem[{Guidorzi {et~al.}(2017)Guidorzi \& Margutti {et~al.}}]{Guidorzi_2017}
Guidorzi, C., Margutti, R., Brout, D., {et~al.} 2017,
  \href{http://dx.doi.org/10.3847/2041-8213/aaa009}{\JournalTitle{The
  Astrophysical Journal}, 851, L36}

\bibitem[{{Hotokezaka} {et~al.}(2019){Hotokezaka} \& {Nakar}
  {et~al.}}]{Hotokezaka}
{Hotokezaka}, K., {Nakar}, E., {Gottlieb}, O., {et~al.} 2019,
  \href{http://dx.doi.org/10.1038/s41550-019-0820-1}{\JournalTitle{Nature
  Astronomy}, 3, 940}, \eprint arXiv:{1806.10596}

\bibitem[{{Howlett} \& {Davis}(2020)}]{howlett_davis}
{Howlett}, C. \& {Davis}, T.~M. 2020,
  \href{http://dx.doi.org/10.1093/mnras/staa049}{\JournalTitle{\mnras}, 492,
  3803}, \eprint arXiv:{1909.00587}

\bibitem[{Howlett {et~al.}(2017)Howlett \& Robotham \& Lagos \&
  Kim}]{Howlett_2017}
Howlett, C., Robotham, A. S.~G., Lagos, C. D.~P., \& Kim, A.~G. 2017,
  \href{http://dx.doi.org/10.3847/1538-4357/aa88c8}{\JournalTitle{The
  Astrophysical Journal}, 847, 128}

\bibitem[{{Howlett} {et~al.}(2015){Howlett} \& {Ross} \& {Samushia} \&
  {Percival} \& {Manera}}]{sdssmgs}
{Howlett}, C., {Ross}, A.~J., {Samushia}, L., {Percival}, W.~J., \& {Manera},
  M. 2015,
  \href{http://dx.doi.org/10.1093/mnras/stu2693}{\JournalTitle{\mnras}, 449,
  848}, \eprint arXiv:{1409.3238}

\bibitem[{{Howlett} {et~al.}(2017{\natexlab{a}}){Howlett} \& {Staveley-Smith}
  \& {Blake}}]{2017MNRAS.464.2517H}
{Howlett}, C., {Staveley-Smith}, L., \& {Blake}, C. 2017{\natexlab{a}},
  \href{http://dx.doi.org/10.1093/mnras/stw2466}{\JournalTitle{\mnras}, 464,
  2517}, \eprint arXiv:{1609.08247}

\bibitem[{{Howlett} {et~al.}(2017{\natexlab{b}}){Howlett} \& {Staveley-Smith}
  {et~al.}}]{Howlett_2017a}
{Howlett}, C., {Staveley-Smith}, L., {Elahi}, P.~J., {et~al.}
  2017{\natexlab{b}},
  \href{http://dx.doi.org/10.1093/mnras/stx1521}{\JournalTitle{\mnras}, 471,
  3135}, \eprint arXiv:{1706.05130}

\bibitem[{{Hui} \& {Greene}(2006)}]{2006PhRvD..73l3526H}
{Hui}, L. \& {Greene}, P.~B. 2006,
  \href{http://dx.doi.org/10.1103/PhysRevD.73.123526}{\JournalTitle{PRD}, 73,
  123526}, \eprint{astro-ph/0512159}

\bibitem[{{Huterer} {et~al.}(2017){Huterer} \& {Shafer} \& {Scolnic} \&
  {Schmidt}}]{huterer}
{Huterer}, D., {Shafer}, D.~L., {Scolnic}, D.~M., \& {Schmidt}, F. 2017,
  \href{http://dx.doi.org/10.1088/1475-7516/2017/05/015}{\JournalTitle{\jcap},
  2017, 015}, \eprint arXiv:{1611.09862}

\bibitem[{Huterer {et~al.}(2015)Huterer \& Kirkby {et~al.}}]{HUTERER201523}
Huterer, D., Kirkby, D., Bean, R., {et~al.} 2015,
  \href{http://dx.doi.org/https://doi.org/10.1016/j.astropartphys.2014.07.004}{\JournalTitle{Astroparticle
  Physics}, 63, 23 }, dark Energy and CMB

\bibitem[{{Icaza-Lizaola} {et~al.}(2020){Icaza-Lizaola} \& {Vargas-Maga{\~n}a}
  {et~al.}}]{eboss}
{Icaza-Lizaola}, M., {Vargas-Maga{\~n}a}, M., {Fromenteau}, S., {et~al.} 2020,
  \href{http://dx.doi.org/10.1093/mnras/stz3602}{\JournalTitle{\mnras}, 492,
  4189}, \eprint arXiv:{1909.07742}

\bibitem[{{Johnston} {et~al.}(2008){Johnston} \& {Taylor}
  {et~al.}}]{2008ExA....22..151J}
{Johnston}, S., {Taylor}, R., {Bailes}, M., {et~al.} 2008,
  \href{http://dx.doi.org/10.1007/s10686-008-9124-7}{\JournalTitle{Experimental
  Astronomy}, 22, 151}, \eprint arXiv:{0810.5187}

\bibitem[{{Kaiser}(1987)}]{1987MNRAS.227....1K}
{Kaiser}, N. 1987,
  \href{http://dx.doi.org/10.1093/mnras/227.1.1}{\JournalTitle{\mnras}, 227, 1}

\bibitem[{{Kawamura} {et~al.}(2006){Kawamura} \& {Nakamura} {et~al.}}]{decigo}
{Kawamura}, S., {Nakamura}, T., {Ando}, M., {et~al.} 2006,
  \href{http://dx.doi.org/10.1088/0264-9381/23/8/S17}{\JournalTitle{Classical
  and Quantum Gravity}, 23, S125}

\bibitem[{Keeley {et~al.}(2019)Keeley \& Shafieloo \& L’Huillier \&
  Linder}]{Keeley_2019}
Keeley, R.~E., Shafieloo, A., L’Huillier, B., \& Linder, E.~V. 2019,
  \href{http://dx.doi.org/10.1093/mnras/stz3304}{\JournalTitle{Monthly Notices
  of the Royal Astronomical Society}, 491, 3983–3989}

\bibitem[{{Kim} {et~al.}(2019){Kim} \& {Aldering} {et~al.}}]{kim_wp}
{Kim}, A., {Aldering}, G., {Antilogus}, P., {et~al.} 2019,
  \JournalTitle{\baas}, 51, 140, \eprint arXiv:{1903.07652}

\bibitem[{{Kim} \& {Linder}(2020)}]{kim2020}
{Kim}, A.~G. \& {Linder}, E.~V. 2020,
  \href{http://dx.doi.org/10.1103/PhysRevD.101.023516}{\JournalTitle{\prd},
  101, 023516}, \eprint arXiv:{1911.09121}

\bibitem[{Lewis \& Bridle(2002)}]{Lewis:2002ah}
Lewis, A. \& Bridle, S. 2002,
  \href{http://dx.doi.org/10.1103/PhysRevD.66.103511}{\JournalTitle{\prd}, 66,
  103511}, \eprint arXiv:{astro-ph/0205436}

\bibitem[{{LIGO Scientific Collaboration} {et~al.}(2017){LIGO Scientific
  Collaboration} \& {Virgo Collaboration} {et~al.}}]{MMApaper}
{LIGO Scientific Collaboration}, {Virgo Collaboration}, {GBM}, F., {et~al.}
  2017, \JournalTitle{ArXiv e-prints}, \eprint arXiv:{1710.05833}

\bibitem[{{Linder}(2005)}]{2005PhRvD..72d3529L}
{Linder}, E.~V. 2005,
  \href{http://dx.doi.org/10.1103/PhysRevD.72.043529}{\JournalTitle{\prd}, 72,
  043529}, \eprint arXiv:{astro-ph/0507263}

\bibitem[{Linder(2020)}]{linder2020limited}
Linder, E.~V. 2020, Limited Modified Gravity, \eprint arXiv:{2003.10453}

\bibitem[{{Linder} \& {Cahn}(2007)}]{2007APh....28..481L}
{Linder}, E.~V. \& {Cahn}, R.~N. 2007,
  \href{http://dx.doi.org/10.1016/j.astropartphys.2007.09.003}{\JournalTitle{Astroparticle
  Physics}, 28, 481}, \eprint arXiv:{astro-ph/0701317}

\bibitem[{{Metzger} \& {Berger}(2012)}]{2012ApJ...746...48M}
{Metzger}, B.~D. \& {Berger}, E. 2012,
  \href{http://dx.doi.org/10.1088/0004-637X/746/1/48}{\JournalTitle{\apj}, 746,
  48}, \eprint arXiv:{1108.6056}

\bibitem[{{Metzger} {et~al.}(2010){Metzger} \& {Mart{\'\i}nez-Pinedo}
  {et~al.}}]{Metzger}
{Metzger}, B.~D., {Mart{\'\i}nez-Pinedo}, G., {Darbha}, S., {et~al.} 2010,
  \href{http://dx.doi.org/10.1111/j.1365-2966.2010.16864.x}{\JournalTitle{\mnras},
  406, 2650}, \eprint arXiv:{1001.5029}

\bibitem[{{Mortlock} {et~al.}(2018){Mortlock} \& {Feeney} \& {Peiris} \&
  {Williamson} \& {Nissanke}}]{2018arXiv181111723M}
{Mortlock}, D.~J., {Feeney}, S.~M., {Peiris}, H.~V., {Williamson}, A.~R., \&
  {Nissanke}, S.~M. 2018, \JournalTitle{arXiv e-prints}, arXiv:1811.11723,
  \eprint arXiv:{1811.11723}

\bibitem[{{Mukherjee} {et~al.}(2019){Mukherjee} \& {Lavaux}
  {et~al.}}]{Mukherjee}
{Mukherjee}, S., {Lavaux}, G., {Bouchet}, F.~R., {et~al.} 2019,
  \JournalTitle{arXiv e-prints}, arXiv:1909.08627, \eprint arXiv:{1909.08627}

\bibitem[{Nicolaou {et~al.}(2019)Nicolaou \& Lahav \& Lemos \& Hartley \&
  Braden}]{nicolaou2019impact}
Nicolaou, C., Lahav, O., Lemos, P., Hartley, W., \& Braden, J. 2019, The Impact
  of Peculiar Velocities on the Estimation of the Hubble Constant from
  Gravitational Wave Standard Sirens, \eprint arXiv:{1909.09609}

\bibitem[{Nishizawa(2018)}]{Nishizawa_2018}
Nishizawa, A. 2018,
  \href{http://dx.doi.org/10.1103/physrevd.97.104037}{\JournalTitle{Physical
  Review D}, 97}

\bibitem[{{Oka} {et~al.}(2014){Oka} \& {Saito} \& {Nishimichi} \& {Taruya} \&
  {Yamamoto}}]{sdsslrg}
{Oka}, A., {Saito}, S., {Nishimichi}, T., {Taruya}, A., \& {Yamamoto}, K. 2014,
  \href{http://dx.doi.org/10.1093/mnras/stu111}{\JournalTitle{\mnras}, 439,
  2515}, \eprint arXiv:{1310.2820}

\bibitem[{{Palmese} {et~al.}(2019){Palmese} \& {Graur} {et~al.}}]{palmese_WP}
{Palmese}, A., {Graur}, O., {Annis}, J.~T., {et~al.} 2019,
  \JournalTitle{\baas}, 51, 310, \eprint arXiv:{1903.04730}

\bibitem[{Punturo {et~al.}(2010)Punturo \& Abernathy {et~al.}}]{Punturo_2010}
Punturo, M., Abernathy, M., Acernese, F., {et~al.} 2010,
  \href{http://dx.doi.org/10.1088/0264-9381/27/19/194002}{\JournalTitle{Classical
  and Quantum Gravity}, 27, 194002}

\bibitem[{Saltas {et~al.}(2014)Saltas \& Sawicki \& Amendola \&
  Kunz}]{Saltas_2014}
Saltas, I.~D., Sawicki, I., Amendola, L., \& Kunz, M. 2014,
  \href{http://dx.doi.org/10.1103/physrevlett.113.191101}{\JournalTitle{Physical
  Review Letters}, 113}

\bibitem[{{Sathyaprakash} {et~al.}(2010){Sathyaprakash} \& {Schutz} \& {Van Den
  Broeck}}]{2010CQGra..27u5006S}
{Sathyaprakash}, B.~S., {Schutz}, B.~F., \& {Van Den Broeck}, C. 2010,
  \href{http://dx.doi.org/10.1088/0264-9381/27/21/215006}{\JournalTitle{Classical
  and Quantum Gravity}, 27, 215006}, \eprint arXiv:{0906.4151}

\bibitem[{{Satpathy} {et~al.}(2017){Satpathy} \& {Alam} {et~al.}}]{boss}
{Satpathy}, S., {Alam}, S., {Ho}, S., {et~al.} 2017,
  \href{http://dx.doi.org/10.1093/mnras/stx883}{\JournalTitle{\mnras}, 469,
  1369}, \eprint arXiv:{1607.03148}

\bibitem[{{Saunders} {et~al.}(1998){Saunders} \& {Oliver} {et~al.}}]{pscz}
{Saunders}, W., {Oliver}, S., {Keeble}, O., {et~al.} 1998, in Wide Field
  Surveys in Cosmology, ed. S.~{Colombi}, Y.~{Mellier}, \& B.~{Raban}, Vol.~14,
  71

\bibitem[{{Schutz}(1986)}]{schutz}
{Schutz}, B.~F. 1986,
  \href{http://dx.doi.org/10.1038/323310a0}{\JournalTitle{\nat}, 323, 310}

\bibitem[{Scolnic {et~al.}(2017)Scolnic \& Kessler {et~al.}}]{Scolnic_2017}
Scolnic, D., Kessler, R., Brout, D., {et~al.} 2017,
  \href{http://dx.doi.org/10.3847/2041-8213/aa9d82}{\JournalTitle{The
  Astrophysical Journal}, 852, L3}

\bibitem[{Shafieloo {et~al.}(2020)Shafieloo \& Keeley \&
  Linder}]{Shafieloo_2020}
Shafieloo, A., Keeley, R.~E., \& Linder, E.~V. 2020,
  \href{http://dx.doi.org/10.1088/1475-7516/2020/03/019}{\JournalTitle{Journal
  of Cosmology and Astroparticle Physics}, 2020, 019–019}

\bibitem[{{Soares-Santos} {et~al.}(2019){Soares-Santos} \& {Palmese}
  {et~al.}}]{darksiren1}
{Soares-Santos}, M., {Palmese}, A., {Hartley}, W., {et~al.} 2019,
  \href{http://dx.doi.org/10.3847/2041-8213/ab14f1}{\JournalTitle{\apjl}, 876,
  L7}, \eprint arXiv:{1901.01540}

\bibitem[{Song \& Percival(2009)}]{Song_2009}
Song, Y.-S. \& Percival, W.~J. 2009,
  \href{http://dx.doi.org/10.1088/1475-7516/2009/10/004}{\JournalTitle{Journal
  of Cosmology and Astroparticle Physics}, 2009, 004–004}

\bibitem[{{Swann} {et~al.}(2019){Swann} \& {Sullivan} {et~al.}}]{4most}
{Swann}, E., {Sullivan}, M., {Carrick}, J., {et~al.} 2019,
  \href{http://dx.doi.org/10.18727/0722-6691/5129}{\JournalTitle{The
  Messenger}, 175, 58}, \eprint arXiv:{1903.02476}

\bibitem[{Takahashi \& Nakamura(2003)}]{decigo2}
Takahashi, R. \& Nakamura, T. 2003,
  \href{http://dx.doi.org/10.1086/379112}{\JournalTitle{The Astrophysical
  Journal}, 596, L231–L234}

\bibitem[{{The LIGO Scientific Collaboration} {et~al.}(2020){The LIGO
  Scientific Collaboration} \& {the Virgo Collaboration} {et~al.}}]{190425}
{The LIGO Scientific Collaboration}, {the Virgo Collaboration}, {Abbott},
  B.~P., {et~al.} 2020, \JournalTitle{arXiv e-prints}, arXiv:2001.01761,
  \eprint arXiv:{2001.01761}

\bibitem[{{Vitale} \& {Chen}(2018)}]{vitale}
{Vitale}, S. \& {Chen}, H.-Y. 2018,
  \href{http://dx.doi.org/10.1103/PhysRevLett.121.021303}{\JournalTitle{\prl},
  121, 021303}, \eprint arXiv:{1804.07337}

\bibitem[{{Wang} {et~al.}(2018){Wang} \& {Wang} \& {Zou}}]{2018PhRvD..98f3503W}
{Wang}, Y.~Y., {Wang}, F.~Y., \& {Zou}, Y.~C. 2018,
  \href{http://dx.doi.org/10.1103/PhysRevD.98.063503}{\JournalTitle{\prd}, 98,
  063503}, \eprint arXiv:{1710.06113}

\bibitem[{Zhao {et~al.}(2011)Zhao \& Van Den~Broeck \& Baskaran \&
  Li}]{Zhao_2011}
Zhao, W., Van Den~Broeck, C., Baskaran, D., \& Li, T. G.~F. 2011,
  \href{http://dx.doi.org/10.1103/physrevd.83.023005}{\JournalTitle{Physical
  Review D}, 83}

\bibitem[{{Zhao} \& {Wen}(2018)}]{zhao}
{Zhao}, W. \& {Wen}, L. 2018,
  \href{http://dx.doi.org/10.1103/PhysRevD.97.064031}{\JournalTitle{\prd}, 97,
  064031}, \eprint arXiv:{1710.05325}

\end{thebibliography}
